\documentclass[aps,eqsecnum,nofootinbib,twocolumn]{revtex4}
\usepackage[dvips]{graphicx}  % for graphics
\def\be{\begin{equation}}
\def\ee{\end{equation}}
\def\bea{\begin{eqnarray}}
\def\eea{\end{eqnarray}}

\newcommand{\eq}[1]{Eq.~(\ref{eq:#1})}
\newcommand{\sect}[1]{Sec.~\ref{sec:#1}}
\newcommand{\fig}[1]{Fig.~\ref{fig:#1}}

\begin{document}

%%%%%%%%%%%%%%%%%%%%%%   Title Page   %%%%%%%%%%%%%%%%%%%%%%%
%
%\rightline{KEK-TH-1133}
%\rightline{hep-ph/0701201}
%
\title{
String theory and quark-gluon plasma
}
\author{Makoto Natsuume}
\email{makoto.natsuume@kek.jp}
\affiliation{Theory Division, Institute of Particle and Nuclear Studies, \\
KEK, High Energy Accelerator Research Organization, Tsukuba, Ibaraki, 305-0801, Japan}
\date{\today}
\begin{abstract}
We review the AdS/CFT description of gauge theory plasmas for non-experts. We discuss the low shear viscosity, jet quenching, and $J/\psi$-suppression, which are three major signatures for the quark-gluon plasma observed at RHIC experiments. Based on invited talks presented at ``Frontiers in the physics of quark-gluon plasma" (July 7-8 2006, RIKEN), ``String theory and quantum field theory" (Sep.\ 12-16 2006, YITP), the Fall Meeting 2006 of the Physical Society of Japan (Sep.\ 20-22 2006, Nara Women's Univ.), ``AdS/CFT and strongly coupled quark matter" (Nov.\ 21-22 2006, CCAST), YKIS2006 (Nov.\ 20-28 2006, YITP), a string theory workshop at Rikkyo Univ. (Dec.\ 26-27 2006).
%and "Strongly coupled quark-gluon plasma: SPS, RHIC and LHC"  (Feb.\ 16-18, Nagoya Univ.).
\end{abstract}
%
%\pacs{11.25.-w, 11.25.Tq, 11.25.Uv}
%
\maketitle

%%%%%%%%%
\section{Introduction}\label{sec:intro}
%%%%%%%%%

%\begin{sloppypar}
In this note, we review the connection between AdS/CFT duality and quark-gluon plasma (QGP) experiments at RHIC (see Ref.~\cite{Yagi:2005yb} for a review of QGP physics). RHIC stands for Relativistic Heavy Ion Collider at Brookhaven National Laboratory. The name ``heavy ion" comes from the fact that it collides heavy ions such as gold nuclei $^{197}$Au instead of usual $e^+ e^-$, $pp$ or $p\bar{p}$. The goal of the experiment is to realize the deconfinement transition and form the quark-gluon plasma. In principle, it should be possible to form QGP if one has high enough temperature or high enough density. However, it is not an easy job to confirm QGP formation because of the following problems:
First, what one observes is not QGP itself but only the by-products after hadoronization, and one has to infer what had happened from the by-products. Second, those secondary particles are mostly strongly-interacting, and the perturbative QCD is not very reliable for the current and near-future experimental temperatures.
%\end{sloppypar}

To resolve these problems, many attempts are made to identify the generic signatures of QGP. Some of the generic signatures discussed to date are as follows:
\begin{enumerate}
\item The elliptic flow which may be the consequence of very low viscosity of QGP
\item The jet quenching
\item $J/\Psi$-suppression
\end{enumerate}
All of these signatures have been discussed in the AdS/CFT duality, so I review recent developments focusing on these phenomena.

% for qft2006
%Let me first remind you of the duality. The original AdS/CFT duality is \cite{Witten:1998qj,Witten:1998zw,Gubser:1998bc} (see Ref.~\cite{Aharony:1999ti} for a review)
%
% AdS<-> CFT
%
%The black hole appears on the right-hand side because a black hole is also a thermal system due to the Hawking radiation. As with the original AdS/CFT duality, this duality can be motivated from the near-horizon limit of the D3-brane.

%%%%%%%%%
\section{A short course on string theory}\label{sec:string}
%%%%%%%%%

%%============
\begin{figure}
\begin{center}
\includegraphics{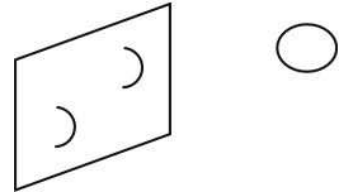}
%\vskip2mm
\caption{Main ingredients of string theory. An open string can have endpoints on the D-brane (left) and describes a gauge theory. A closed string represents a graviton (right).}
\label{fig:string}
\end{center}
\end{figure}
%%============

%%============
\begin{figure}
\begin{center}
\includegraphics{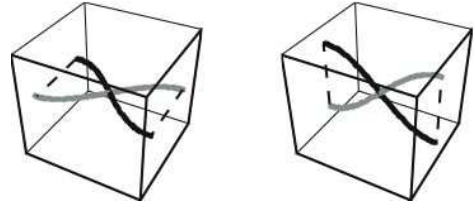}
%\vskip2mm
\caption{The simplest oscillations of an open string.}
\label{fig:level1}
\end{center}
\end{figure}
%%============

%%============
\begin{figure}
\begin{center}
\includegraphics{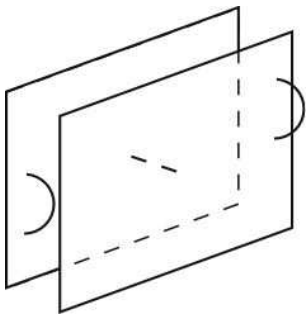}
%\vskip2mm
\caption{$N$ D-branes represent a $SU(N)$ gauge theory.}
\label{fig:SYM}
\end{center}
\end{figure}
%%============

Since this review is aimed at non-experts, I first give a short course on string theory, in particular emphasizing on how gauge theories and black holes are described in string theory.

Figure~\ref{fig:string} shows main ingredients of string theory. There are two kinds of string: open strings with endpoints and closed strings with no endpoints. As I will show you shortly, an open string represents a gauge theory. A closed string represents a graviton. An open string propagates through spacetime just like a closed string, but an open string can also have its endpoints on an object, the so-called D-brane. The open string I consider in this lecture is always of this type.

The easiest way to see that an open string represents a gauge theory is look at  how the string oscillates. Figure~\ref{fig:level1} shows the simplest open string oscillation (in 4 dimensions).%
%YKIS
\footnote{String theory is actually a 10-dimensional theory, but we discuss in 4-dimensions for illustrating purpose.} 
As you can see, the string can oscillate in two directions. So, the open string has two degrees of freedom at this level. These degrees of freedom represent the two polarizations of the photon. In this sense, the open string represents a gauge theory. 

Of course, our interest here is not QED, but rather QCD, so how can one describe a Yang-Mills theory? An open string has endpoints on a D-brane, but if there are $N$ coincident D-branes, open strings can have endpoints in various ways (\fig{SYM}). These new degrees of freedom precisely correspond to $SU(N)$ degrees of freedom. 

%\begin{sloppypar}
Since these open strings are constrained to have their endpoints on the D-brane, the gauge theory described by the D-brane is localized on the D-brane. The D-branes arise with various dimensionalities. A D-brane with a $p$-dimensional spatial extension is called the D$p$-brane. Namely, the D0-brane is point-like, the D1-brane is string-like, the D2-brane is membrane-like, and so on. 
Thus, the D$p$-brane describes a $(p+1)$-dimensional Yang-Mills theory. We are interested in 4-dimensional gauge theories, so consider the D3-brane in order to mimic QCD.
%\end{sloppypar}

\begin{figure}
\begin{center}
\includegraphics{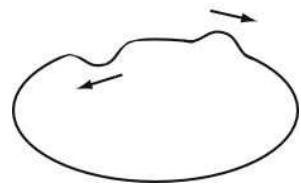}
%\vskip2mm
\caption{A closed string has two independent modes, which corresponds to the spin-2 nature of the graviton.}
\label{fig:closed}
\end{center}
\end{figure}

On the other hand, a closed string represents a graviton. Again, the easiest way to see this is to look at how the string oscillates (\fig{closed}). In general, the oscillations on a string have two modes: the left-moving modes and right-moving modes. For an open string, these modes mix each other at endpoints, but these modes become independent for a closed string. So, one can oscillate the right-moving mode in one direction and the left-moving mode in the other direction. In a sense, a closed string oscillates in two directions simultaneously. This property explains the spin-2 nature of the graviton. In fact, a graviton also oscillates in two directions simultaneously. Strictly speaking, a closed string represents a graviton and two undiscovered scalar particles, the dilaton and the axion. Since each mode has two degrees of freedom, a closed string has 4 degrees of freedom at this level (in 4 dimensions). The graviton has only 2 degrees of freedom, and two scalar fields cover the remaining degrees of freedom.

%%============
%%============

%\begin{sloppypar}
We have seen that we can get a gauge theory from the D-brane, but it is not clear if the D-brane is simply described by a gauge theory. This is because string theory is more than a gauge theory, namely it has gravitons. At this point, it is not clear if the effect of gravity can be neglected. According to general relativity, any energy-momentum tensor curves spacetime. The D-brane of course has some energy, so how the D-branes curve spacetime? Since gravity is described by the Newton potential
\be
\phi_{\rm Newton} \sim \frac{GM}{r}~,
\ee
one can measure the effect of curvature by $GM$. According to string theory, the Newton constant $G$ and the mass $M$ of the D-brane are given by
\bea
G &\sim& g_s^2~, \\
M &\sim& N/g_s~,
\eea
where $g_s$ is the string coupling constant which governs the strength of the interactions between strings.% 
\footnote{Note that the D-brane is infinitely heavy in the weak coupling limit $g_s \ll 1$, which is consistent with the hypersurface picture of the D-brane, and the open strings attached represent the fluctuations of the D-brane.} 
So, one gets $GM \sim g_s N$. This means that as long as $g_s N \ll 1$, one can neglect the effects of gravity and spacetime remains flat. In this case, the D-brane is simply described by a gauge theory.
%\end{sloppypar}

On the other hand, when $g_s N \gg 1$, the D-brane starts to curve spacetime, and eventually it should become a black hole. So, in this case, the D-brane can be described by a black hole. The black hole here is not the usual Schwarzschild-like black hole. We consider a D-brane, an object with infinite spatial extension. Thus, the black hole formed from the D-brane has an horizon which extends indefinitely and is called the ``black brane."

%%============
\begin{figure}
\begin{center}
\includegraphics{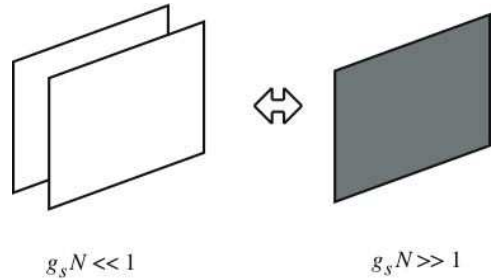}
%\vskip2mm
\caption{A gauge theory is described by a black hole in the strong coupling.}
\label{fig:largeN}
\end{center}
\end{figure}
%%============

To summarize, one can describe the D-brane both by a gauge theory and by a black hole; which description is better depends on the value of $g_s N$. The string coupling and the Yang-Mills coupling $g_{\rm YM}$ are related by $g_s \sim g_{\rm YM}^2$ 
%YKIS
(see \sect{details}), 
so $g_s N$ is nothing but the standard `t~Hooft coupling, $\lambda := g_{\rm YM}^2 N$. This suggests that the black hole is the large `t~Hooft coupling limit of a gauge theory. Since `t~Hooft coupling is the effective coupling of a gauge theory, the black hole is the strong coupling limit of the gauge theory (\fig{largeN}). Thus, the strategy here is to use black holes in order to compute gauge theory observables in the strong coupling regime.%
\footnote{Conversely, such a correspondence has also been used for the microscopic derivation of black hole entropy \cite{Strominger:1996sh}.}

Our argument here is very rough, but more refined version of the argument is known as the AdS/CFT duality \cite{Maldacena:1997re,Witten:1998qj,Witten:1998zw,Gubser:1998bc} (see Ref.~\cite{Aharony:1999ti} for a review). The precise correspondence is as follows:
%\bea
%{\cal N}=4 \mbox{ SYM} 
%&\leftrightarrow& \mbox{type IIB string theory} \nonumber\\
%&& \mbox{on AdS}_5 \times S^5.
%\label{eq:dual}
%\eea
%\bea
%\lefteqn{{\cal N}=4 \mbox{ super-Yang-Mills (SYM)} \leftrightarrow \hspace{4cm}}
%\nonumber\\
%&& \mbox{type IIB string theory} \nonumber\\
%&& \mbox{on AdS}_5 \times S^5.
%\label{eq:dual}
%\eea
\bea
\lefteqn{{\cal N}=4 \mbox{ super-Yang-Mills theory (SYM)} \leftrightarrow}
\nonumber\\
&& \mbox{type IIB string theory on AdS}_5 \times S^5.
\label{eq:dual}
\eea
%\begin{sloppypar}
\noindent
Here, ${\cal N}=4$ means that the theory has 4 supercharges which are the maximum number of supercharges for a 4-dimensional theory. Also, ${\rm AdS}_5$ stands for the five-dimensional anti-deSitter space. DeSitter was a Dutch astronomer who found a solution of Einstein equation with a constant positive curvature in 1917. The space ${\rm AdS}_5$ instead has a constant negative curvature; this explains the prefix ``anti." One can reach this correspondence by studying the D3-brane more carefully, but I will skip the argument. Instead, I explain the correspondence from the symmetry point of view in Appendix.
%YKIS
%Instead, I explain why the geometry ${\rm AdS}_5 \times S^5$ is relevant from the symmetry point of view.
%\end{sloppypar}

We will use the finite temperature version of the duality and its cousins; in this case, one needs to replace ${\rm AdS}_5$ by a black hole in ${\rm AdS}_5$, which is known as the Schwarzschild-${\rm AdS}_5$ black hole:
%\begin{center}
%${\cal N}=4$ SYM at finite temperature 
%$\leftrightarrow$ 
%type IIB string theory in Schwarzschild-${\rm AdS}_5$ black holes (${\rm SAdS}_5$) $\times S^5$
%\end{center}
\bea
\lefteqn{{\cal N}=4 \mbox{ SYM at finite temperature} \leftrightarrow}
\nonumber\\
&& \mbox{type IIB string theory} \nonumber\\
&& \mbox{on (Schwarzschild-AdS}_5 \mbox{ black holes}) \times S^5.
%&& ({\rm SAdS}_5) \times S^5.
\nonumber
\eea

%YKIS
%The gauge theory side is in this case a finite temperature system, but a black hole is also a finite temperature system; a black hole has a temperature due to the Hawking radiation.

%%%%%%%%%
\section{Black holes and hydrodynamics}\label{sec:hydrodynamics}
%%%%%%%%%

%%============
\begin{figure*}
\begin{center}
\includegraphics{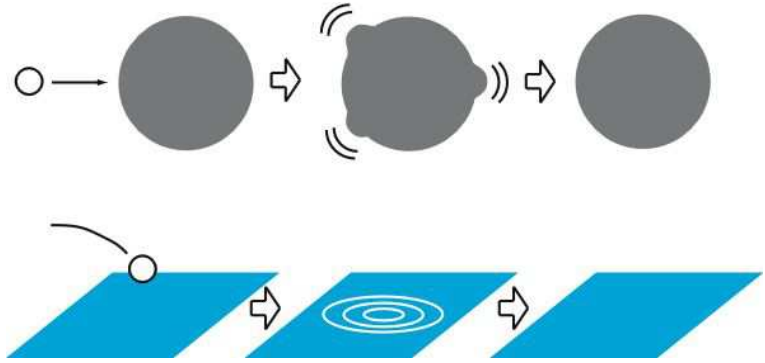}
%\vskip2mm
\caption{When one adds a perturbation to a black hole, the black hole behavior is similar to a hydrodynamic system. In hydrodynamics, this is a consequence of  viscosity.}
\label{fig:hydro}
\end{center}
\end{figure*}
%%============

According to the RHIC experiment, QGP behaves like a liquid. The AdS/CFT then implies that a black hole also behaves like a liquid. Then, plasma quantities should be calculable from black holes. In fact, black holes and hydrodynamic systems behave similarly. Consider adding a perturbation to a black hole, {\it e.g.}, drop some object (\fig{hydro}). Then, the shape of the black hole horizon becomes irregular, but such a perturbation decays quickly, and the black hole returns to the original symmetric shape. The no-hair theorem is one way to see this. According to the theorem, the stationary black hole is unique and symmetric. Thus, the perturbed black hole cannot be stable. If you regard this as a diffusion, the diffusion occurs since the perturbation is absorbed by the black hole.

This behavior is very similar to a liquid. Suppose that one drops a ball in a water pond. 
Then, you generate surface waves, but they decay quickly, and the water pond returns to a state of stable equilibrium. 
%Then, you have waves on the surface, but they decay quickly, and you get the original static surface. 
In hydrodynamics, this is a consequence of viscosity. Thus, one can consider the notion of viscosity for black holes as well. And the ``viscosity" for black holes should be calculable by considering the above process. 

\begin{figure}
\begin{center}
\includegraphics{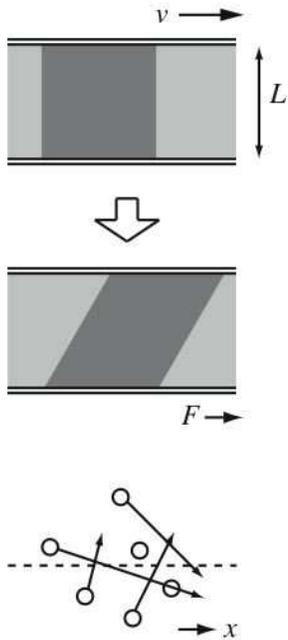}
%\vskip2mm
\caption{When one moves the upper plate, the fluid is dragged due to the viscosity and the lower plate experiences a force. The bottom figure is the close-up view of the fluid.}
\label{fig:viscosity}
\end{center}
\end{figure}

Let me remind you of freshman physics of viscosity. As a simple example, consider a fluid between two plates and move the upper plate with velocity $v$ (\fig{viscosity}). As the fluid is dragged, the lower plate experiences a force. This force is the manifestation of the viscosity. In this case, the force $F$ the lower plate experiences per unit area is given by
\be
\frac{F}{A}=\eta \frac{v}{L}~.
\ee
The proportionality constant $\eta$ is called the (shear) viscosity.

%\begin{sloppypar}
Microscopically, the viscosity arises due to the momentum transfer between molecules. Figure~\ref{fig:viscosity} shows a close-up view of the fluid and I put an artificial boundary to divide the fluid into two parts. The molecules collide with each other and are exchanged randomly through the boundary. But in the situation where you move the upper plate, the molecules in the upper-half part, on average, have more momentum in the $x$-direction than the ones in the lower-half part. These molecules are exchanged, which means that momentum in the $x$-direction is transported through the boundary.
%\end{sloppypar}

Going back to the black hole, how can one calculate the plasma viscosity? I will first give a brief explanation and its implications and then I will justify the claim more in detail.

%%============
%%============

%%===========================================================
\subsection{Quick argument}\label{sec:quick}
%%===========================================================

There are many ways to compute the viscosity and I will explain one simple method. % To do so, let me go back to the D3-brane from the SAdS. 
In the gravity side, the diffusion occurs by black hole absorption. So, it is natural to associate the shear viscosity with the absorption cross section by black holes. (I explain this point more in detail later. The detailed argument suggests that this is the cross section for the graviton of particular polarization.) Now, there is a general theorem on black holes \cite{Das:1996we} that states that the cross section $\sigma_{\rm BH}$ is equal to the horizon area $A$ for a broad range of black holes%
\footnote{As a simple example, this is true for the usual Schwarzschild black hole as well. Precisely speaking, the theorem applies only to the low energy limit $\omega \rightarrow 0$.}
:
\be
\eta \propto \lim_{\omega \rightarrow 0}\sigma_{\rm BH}=A~.
\label{eq:absorption}
\ee
But the horizon area is the famous quantity, namely it represents the black hole entropy, so it must be the plasma entropy
\footnote{We use the unit $c=1$ in this lecture.}
\be
S_{\rm BH} = \frac{A}{4G\hbar} k_{\rm B}
\label{eq:entropy}
\ee
($k_{\rm B}$: Boltzmann constant). Then, the shear viscosity divided by the entropy becomes constant.%
\footnote{Precisely speaking, one divides by the entropy density $s$. We consider black holes with infinite extension, so the entropy itself diverges. The area for the absorption cross section should be understood in a similar way.}
The constant can be determined from the argument later and the result is
\be
\frac{\eta}{s} =\frac{\hbar}{4\pi k_{\rm B}}~.
\label{eq:viscosity_bound}
\ee
This value is very small. In comparison, $\eta/s$ for water is about $3 \times 10^3$ under normal circumstances.

Now, the point is that all the relations we used (cross section versus horizon area, black hole entropy versus horizon area) are generic, so the result must be universal as well. Namely, it does not depend on the details of black holes nor the details of gauge theories. So, the claim is \cite{Kovtun:2004de}
%revtex
\begin{center}
{\it Gauge theory plasmas which have gravity duals have a universal low value of $\eta/s$ 
at large 't~Hooft coupling.
% (at zero chemical potential).
}%
%\footnote{The assumption of zero chemical potential arises from a somewhat technical assumption I did not mention.}
\end{center}
This is a rather indirect argument, but this claim has indeed been checked for many known gravity duals.

\begin{figure}
\begin{center}
\includegraphics{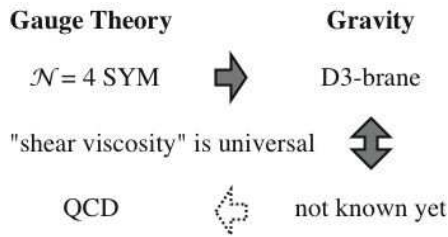}
%\vskip2mm
\caption{Gauge theories described by D-branes are mostly supersymmetric gauge  theories, and not the real QCD. However, the black hole quantity corresponding to the ``shear viscosity" is universal, so probably the results for supersymmetric gauge theories are directly applicable to the real QCD.}
\label{fig:universality}
\end{center}
\end{figure}

%\begin{sloppypar}
This result is very important, so let me rephrase in a different way (\fig{universality}). Gauge theories of which we can actually compute the shear viscosity are supersymmetric gauge theories, not the real QCD. We compute the shear viscosity from black holes, but the gravity dual of QCD is not known. So, one cannot use AdS/CFT directly to compute QCD properties. However, as we saw, the quantity corresponding to the shear viscosity is universal on the black hole side, so one can immediately apply the ${\cal N}=4$ result to the real QCD even though these two theories are completely different.
%\end{sloppypar}

%%============
%\begin{wrapfigure}{r}{9cm}
%\end{wrapfigure}
%%============
In fact, RHIC suggests that QGP has a very low viscosity and the estimated value \cite{Teaney:2003kp,Hirano:2005wx}
\be
\frac{\eta}{s} \sim O(0.1) \times \frac{\hbar}{k_{\rm B}} ?
\ee
is very close to the above AdS/CFT value.%
\footnote{See Refs.~\cite{Gavin:2006xd,Adare:2006nq} for recent estimates.}
 One important point is that the temperature in question is still order of $\Lambda_{\rm QCD}$. At this range of temperature, QCD is still strongly coupled and pQCD is not very reliable. In fact, the naive extrapolation of the weak coupling result gives a larger value for $\eta/s$. Thus, the AdS/CFT duality which predicts the strong coupling behavior may be useful to analyze QGP.

% comment out for hep-ph
%Let me briefly mention the weak coupling behavior. In general, the shear viscosity is given by
%\be
%\eta \sim \rho \bar{v}l_{\rm mfp}
%\ee
%($\rho$: mass density, $\bar{v}$: mean velocity, $l_{\rm mfp}$: mean free path).
%At weak coupling, the interaction becomes weaker, so the mean free path becomes larger. The viscosity arises due to the momentum transfer, and the transfer is more effective at weak coupling. So, at weak coupling, one expects $\eta/s \gg 1/(4\pi)$. Figure~XX summarizes $\eta/s$ both at weak coupling (perturbative ${\cal N}=4$) and at strong coupling (AdS/CFT). As one can see, these two behaviors are very different.

%%===========================================================
\subsection{More in detail}\label{sec:details}
%%===========================================================

%%============
\begin{figure}[tb]
\begin{center}
%\begin{wrapfigure}{r}{7cm}
\includegraphics{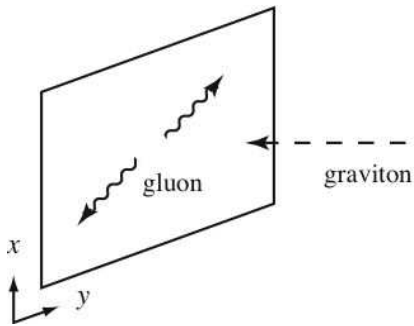}
%\vskip2mm
\caption{The bulk graviton produces the back-to-back scattering of gluons on the boundary. In black hole picture, it is natural to regard the graviton decay rate as the absorption cross section by the black hole.}
\label{fig:absorption}
%\end{wrapfigure}
\end{center}
\end{figure}
%%============
I now describe the relation between the shear viscosity and the absorption cross section. To do so, one first has to understand the interactions of bulk and boundary fields.
% string interactions
These fields can interact at the boundary. 
For example, the graviton produces the back-to-back scattering of gluons (\fig{absorption}).
The relevant interactions are
% Gravity (closed string) couples to any energy-momentum tensor, and the gauge fields (open string) is not an exception.
% T_{xy}->T_{xy}^{\rm YM}
% T_{\mu\nu}->T_{\mu\nu}^{\rm YM}
\be
S_{\rm int} \sim \int d^4x\, \delta\phi F_{\mu\nu}^2 + \delta h^{\mu\nu} T_{\mu\nu}^{\rm YM} + \cdots
\label{eq:interactions}
\ee
($\phi$: dilaton, $h^{\mu\nu}$: graviton, $F_{\mu\nu}$ and $T_{\mu\nu}^{\rm YM}$: the field strength and the energy-momentum tensor of the gauge theory). Such interactions can be obtained by expanding the D-brane action around the expectation values of the bulk fields. The D-brane is described by the so-called Dirac-Born-Infeld (DBI) action. The DBI-action with at most two derivatives contains the following term:
\be
{\cal L} \sim e^{-\phi} F_{\mu\nu}^2 +\cdots~.
\ee
If $\phi$ is constant, one gets the standard gauge theory Lagrangian with $g_{\rm YM}^2 \sim g_s$, where $g_s:=e^{\phi}$. If $\phi$ fluctuates, the action is expanded as
\be
{\cal L} \sim e^{-\langle\phi\rangle} 
\left( F_{\mu\nu}^2 +\delta\phi F_{\mu\nu}^2 \right)~,
\ee
where $\phi=\langle\phi\rangle-\delta\phi$. The second term is nothing but the first term in \eq{interactions}.

This fact could be stated as 
\begin{center}
{\it Bulk field fluctuations act as sources of boundary fields}
\end{center}
I just rephrase the same statement, but this leads to the so-called ``GKP-Witten relation," which is the definition of the AdS/CFT.

Given the interaction term, one can easily calculate the graviton decay rate (with polarization $h_{xy}$, where $x$ and $y$ are directions along the brane) from the standard field theory formula:
%revtex
\bea
\sigma_{\rm QFT}
&=& \frac{1}{S_f}\frac{1}{2\omega} \sum_{\mbox{\scriptsize final states}} \int 
\frac{d^3 p_1}{(2\pi)^3 2\omega_1}\frac{d^3 p_2}{(2\pi)^3 2\omega_2} 
\nonumber\\
&&\mbox{} \times (2\pi)^4 \delta^4(p_f-p_i) |{\cal M}|^2 
\nonumber\\
&=& \frac{8\pi G}{\hbar\omega}\! \int d^4 x \, e^{i\omega t} \langle [T_{xy}^{\rm YM}(t,x),T_{xy}^{\rm YM}(0,0)] \rangle~.
\nonumber
\eea
The first equality is just the Fermi's golden rule with matrix element ${\cal M}$. The matrix element is proportional to $T_{xy}^{\rm YM}$, so the formula is written by a correlator of the energy-momentum tensor. (This is an optical theorem.) The factor $S_f$ is a statistical factor for identical particles in the final state.

In black hole description, it is natural to regard this decay rate as the absorption cross section of the graviton. This has been checked for the D3-brane at zero temperature \cite{Klebanov:1997kc}, so %
\footnote{As a cautionary remark, one should note that \eq{correlator} is known to hold for the D3-brane at zero temperature but it is unclear if it also holds for generic AdS black holes. The problem is that a black hole is the strong coupling limit of a gauge theory, so \eq{correlator} is the strong coupling statement. But, of course, it is not easy to compute the right-hand side in gauge theory. One can compute it in the weak coupling, but weak coupling results differ from strong coupling results in general as we will see in \sect{others}. In the special case of the zero-temperature D3-brane, such a comparison actually makes sense due to the nonrenormalization theorem, and this is the case where \eq{correlator} has been checked. But it is an open question if this is also true for generic AdS black holes. Here, we simply assume that the relation holds at strong coupling.}
\be
\sigma_{\rm BH}=\frac{8\pi G}{\hbar\omega}\! \int d^4 x \, e^{i\omega t} \langle [T_{xy}^{\rm YM}(t,x),T_{xy}^{\rm YM}(0,0)] \rangle~.
\label{eq:correlator}
\ee
We can use this relation to compute the shear viscosity since the viscosity is given by a Kubo formula microscopically: 
\be
\eta = \lim_{\omega \rightarrow 0}
\frac{1}{2\hbar\omega} \!\int d^4 x \, e^{i\omega t} \langle [T_{xy}^{\rm YM}(t,x),T_{xy}^{\rm YM}(0,0)] \rangle~.
\ee
This formula has the same form as the absorption cross section, so one can immediately write the viscosity in terms of the cross section:
\be
\eta = \frac{\lim_{\omega \rightarrow 0}\sigma_{\rm BH}}{16 \pi G}~.
\ee
We have written plasma quantities in terms of black hole quantities. Using Eqs.~(\ref{eq:absorption}) and (\ref{eq:entropy}), one obtains
\be
\frac{\eta}{s} = \frac{\frac{A}{16 \pi G}}{\frac{A}{4G\hbar} k_{\rm B}}
=\frac{\hbar}{4\pi k_{\rm B}}~.
\ee
This is the previous formula (\ref{eq:viscosity_bound}).
%\footnote{To be clear, $G$ is the 10-dimensional Newton constant and $A$ is the horizon area in 7-dimensional spacetime, namely 5-dimensional area (since it appears in the entropy density.)}

%%===========================================================
\subsection{Further checks of universality}\label{sec:chemical}
%%===========================================================

%We saw that gauge theories at strong coupling have a universality of shear viscosity. However, there is an important restriction. There are many proofs of the universality, but all fail in the presence of a chemical potential \cite{Kovtun:2003wp,Buchel:2003tz,Kovtun:2004de,Buchel:2004qq}. So, the natural question is what happens to the universality at finite chemical potential.

We saw that in gauge theories at strong coupling the ratio of shear viscosity to entropy density is universal. How far has the universality been shown? Recently, the universality has been extended to a variety of situations which are not covered in the original proofs. 

The shear viscosity was first computed for the D3-brane \cite{Policastro:2001yc}. This corresponds to the ${\cal N}=4$ SYM which is a scale-invariant theory. (Actually, the theory has a larger symmetry, a conformal symmetry. See Appendix.) But the universality has been shown even for theories with scales, {\it i.e.}, nonconformal theories. Examples include D$p$-branes for $p \neq 3$, the  Klebanov-Tseytlin geometry, the Maldacena-Nunez geometry, and the ${\cal N}=2^*$ system (These theories also have a reduced number of supersymmetries). In fact, various universality arguments were demonstrated for these cases \cite{Kovtun:2004de,Kovtun:2003wp,Buchel:2003tz,Buchel:2004qq}. 

%\begin{sloppypar}
However, these proofs did not cover the cases with a chemical potential. So, the natural question is what happens to the universality at finite chemical potential.
Actually, it is not easy to realize the realistic finite density in AdS/CFT, {\it i.e.}, baryon number density. But there is a simple alternative. Namely, consider charged AdS black holes instead of neutral black holes. A black hole is known to obey thermodynamic-like laws and its first law is written as
\be
dM = TdS_{\rm BH}+\Phi dQ
\ee
($T$: black hole temperature, $\Phi$: electromagnetic potential, $Q$: black hole charge). As one can see, the electromagnetic field $\Phi$ plays a role of a chemical potential.
%\end{sloppypar}

In AdS/CFT, such a charge arises as follows. As in \eq{dual}, the full geometry  involves $S^5$. One can add an angular momentum along $S^5$, which is known as the ``spinning" D3-brane solutions \cite{Kraus:1998hv,Cvetic:1999xp}. The angular momentum becomes a Kaluza-Klein charge after the $S^5$ reduction. 

What is the gauge theory interpretation of the charge? The symmetry of $S^5$ corresponds to an internal symmetry of the SYM, R-symmetry $SO(6)$. This $SO(6)$ rotates adjoint scalars in the ${\cal N}=4$ supermultiplet. (See Appendix.)
% (These scalars correspond to string oscillations in the transverse directions to the brane as described in Appendix. Thus, there are $D-(p+1)=6$ degrees of freedom, and they label $S^5$.) 
 The R-symmetry group $SO(6)$ is rank 3, so one can add at most three independent charges. The three-charge solution is known as the STU solution \cite{Behrndt:1998jd}. When all charges are equal, the STU solution is the well-known Reissner-Nordstr\"{o}m-${\rm AdS}_5$ black hole. Thus, the charge in question is a $U(1)_R$ charge.

%\begin{sloppypar}
Because the charge corresponds to the $U(1)_R$ charge, this is by no means realistic. However, the theory has interesting features which are common to the real QCD. For instance, the phase diagram is qualitatively similar to the QCD diagram \cite{Chamblin:1999tk,Cvetic:1999ne}.%
\footnote{This is the case of the RN-AdS black hole with compact horizon or compact SYM. In this review, we consider the RN-AdS black hole with noncompact horizon, which is always in plasma phase.}
This system does not represent a realistic chemical potential, but it may mimic the realistic case and one may learn an interesting lesson for gauge theory plasmas at finite density.
%\end{sloppypar}

The shear viscosity for charged AdS black holes was computed by 4 groups, and the result turns out to be $\eta/s=1/(4\pi)$ again \cite{Mas:2006dy}-\cite{Maeda:2006by}. So, the universality seems to hold even at nonzero chemical potential. If this is true for generic chemical potential, $\eta/s=1/(4\pi)$ may be true even at finite baryon number density. In fact, when there is a chemical potential associated with a global symmetry, the universality has been proven in more general settings as well \cite{Buchel:2006gb}-\cite{Gauntlett:2006ai}.

The theories described so far all have matter in the adjoint representation and not in the fundamental representation such as quarks. However, the universality has been proven even in the presence of fundamental matter \cite{Mateos:2006yd}. One way to include fundamental matter is to include D7-branes in addition to D3-branes. I will describe one simple way to realize fundamental matter in the next section, and the D3-D7 system is an extension of the method. This D3-D7 system is known as Karch-Katz model \cite{Karch:2002sh}.

%\begin{sloppypar}
Gauge theory plasmas described so far are stationary ones. The real plasma at RHIC is of course a rapidly changing system, and it is desirable to study such a plasma as well. In the gravity side, this corresponds to a time-dependent black hole. The universality has been claimed even for such a case \cite{Janik:2006ft}.%
\footnote{In this subsection, we included relatively recent discussions only. See Refs.~\cite{Son:2002sd}-\cite{Buchel:2005cv} for early computations and computations of the other transport coefficients.}
%\end{sloppypar}

%%===========================================================
\subsection{Other issues}\label{sec:others}
%%===========================================================

% comment out for hep-ph
%Let me briefly mention the weak coupling behavior. In general, the shear viscosity is given by
%\be
%\eta \sim \rho \bar{v}l_{\rm mfp}
%\ee
%($\rho$: mass density, $\bar{v}$: mean velocity, $l_{\rm mfp}$: mean free path).
%At weak coupling, the interaction becomes weaker, so the mean free path becomes larger. The viscosity arises due to the momentum transfer, and the transfer is more effective at weak coupling. So, at weak coupling, one expects $\eta/s \gg 1/(4\pi)$. Figure~XX summarizes $\eta/s$ both at weak coupling (perturbative ${\cal N}=4$) and at strong coupling (AdS/CFT). As one can see, these two behaviors are very different.

We saw that $\eta/s=1/(4\pi)$, but this is in the strong coupling limit or in  the large `t~Hooft coupling limit $\lambda\rightarrow\infty$. Real QCD of course has finite coupling, so how about $\eta/s$ at finite coupling? In general, the shear viscosity is given by
\be
\eta \sim \rho \bar{v}l_{\rm mfp}
\ee
($\rho$: mass density, $\bar{v}$: mean velocity, $l_{\rm mfp}$: mean free path).
At weak coupling, the interaction becomes weaker, so the mean free path becomes larger. The viscosity arises due to the momentum transfer, and the transfer is more effective at weak coupling. So, at weak coupling, one expects $\eta/s \gg 1/(4\pi)$.%
\footnote{Actually, we have two parameters $g_{\rm YM}$ and $N$ (or $g_{\rm YM}$ and $\lambda$), so there are two kinds of finite-coupling corrections. First is the finite-$\lambda$ correction described here. This corresponds to the so-called $\alpha'$-corrections in the gravity side. The effective actions of string theory contain higher curvature terms in addition to the Einstein-Hilbert action; {\it e.g.}, $S \sim \int \sqrt{-g} (R+\alpha'^3 O(R^4)+\cdots)$. The finite-$\lambda$ correction comes from such higher curvature terms. On the other hand, the finite-$g_{\rm YM}$ correction comes from the so-called string loop corrections since the string coupling constant is proportional to $g_{\rm YM}^2$. Such a correction to $\eta$ has never been evaluated.}

For the ${\cal N}=4$ SYM, the finite-$\lambda$ correction can be computed, and the correction indeed raises the value of $\eta/s$ \cite{Buchel:2004di}. The correction is about 9\% if one uses $\lambda=6\pi$ (or $\alpha_{\rm SYM}=1/2$). Thus, the strong coupling result seems a good approximation. The universality does not hold for finite-$\lambda$ corrections though, so it is not clear how realistic the result is. 

\begin{figure}[tb]
\begin{center}
\includegraphics{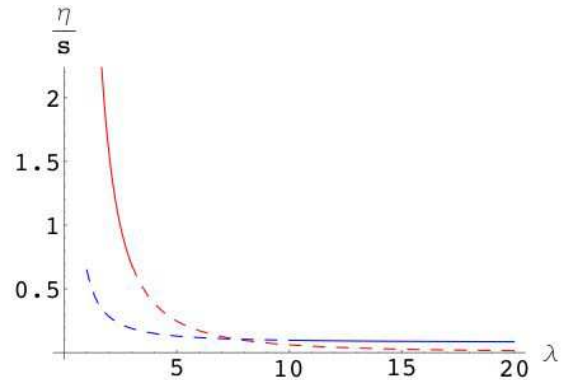}
%\vskip2mm
\caption{The coupling constant dependence of $\eta/s$ both from the weak coupling result and the strong coupling result. The red line represents the weak coupling result, {\it i.e.}, a naive extrapolation of the perturbative ${\cal N}=4$ SYM result (Based on Ref.~\cite{Huot:2006ys}). The blue line represents the strong coupling result, {\it i.e.}, the ${\cal N}=4$ AdS/CFT result with the first order correction in $\lambda$ (Based on Ref.~\cite{Buchel:2004di}).  }
\label{fig:finite_coupling}
\end{center}
\end{figure}

Figure~\ref{fig:finite_coupling} shows both the weak coupling result and the strong coupling result. They both increase at weak coupling. However, there is a difference. The viscosity is proportional to the mean free path, so is inversely proportional to the coupling constant. Thus, the naive extrapolation of the weak coupling result suggests the perfect-fluid-like behavior in the strong coupling limit. Instead, AdS/CFT tells that $\eta/s$ cannot be small indefinitely and is saturated.

Incidentally, our problem is the strong coupling problem, so one might wonder how useful lattice computations are. A lattice computation of the viscosity for the pure Yang-Mills theory has been done in Ref.~\cite{Nakamura:2004sy}. Unfortunately, the result has very large error bars, so one cannot make a definite statement, but it would be very interesting to make the errors smaller.  

%%============
%%============

%%%%%%%%%
\section{Heavy quarks in medium}\label{sec:heavy_quark}
%%%%%%%%%

Recently, there has been much discussion on heavy quark dynamics in plasma medium. Two applications have been discussed: issues related to $J/\psi$ suppression and issues related to jet quenching. In this section, I quickly summarize the discussion. 

%%===========================================================
\subsection{$J/\Psi$-suppression}\label{sec:JPsi}
%%===========================================================

%The next natural step is to investigate $J/\Psi$-suppression. 
Since $J/\Psi$ is heavy, charm pair production occurs only at the early stages of the nuclear collision. However, if the production occurs in the plasma medium, charmonium formation is suppressed due to the Debye screening. One technical difficulty is that the $c\bar{c}$ pair is not produced at rest relative to the plasma. Therefore, the screening length is expected to be  velocity-dependent. Such a computation has been done only for the Abelian plasma \cite{Chu:1988wh}.

%%============
\begin{figure}[tb]
\begin{center}
%\begin{wrapfigure}{r}{10cm}
\includegraphics{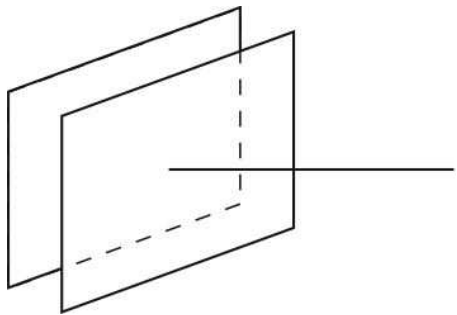}
%\vskip2mm
\caption{An infinitely long string transforms as the fundamental representation.}
\label{fig:quark}
%\end{wrapfigure}
\end{center}
\end{figure}
%%============

%%============
\begin{figure}[tb]
\begin{center}
%\begin{wrapfigure}{r}{10cm}
\includegraphics{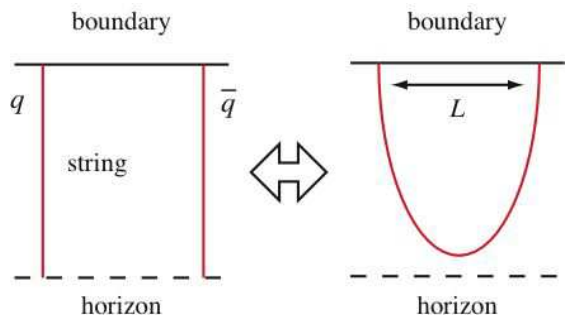}
%\vskip2mm
\caption{For a $q\bar{q}$ pair, the left configuration is not the lowest energy configuration. Instead, the right configuration is energetically favorable. The energy difference is interpreted as a $q\bar{q}$ potential. At finite temperature, this is no longer true. For large enough separation of the pair, the left configuration becomes favorable. This phenomenon is the AdS/CFT description of the Debye screening. }
\label{fig:wilson}
%\end{wrapfigure}
\end{center}
\end{figure}
%%============

% for hep-ph
First, one has to understand how to realize a heavy quark in AdS/CFT. To do so, let us go back to \fig{SYM}. When there are multiple number of D-branes, an open string can have endpoints in various ways; there are $N^2$ possibilities. This means that the string transforms as the adjoint representation of $SU(N)$ gauge theory. Now, consider an infinitely long string (\fig{quark}). In this case, the string can have endpoints in $N$ different ways. This means that the string transforms as the fundamental representation of $SU(N)$ gauge theory. In this sense, such a long string represents a ``quark." Such a string has an extension and tension, so the string has a large mass, which means that the long string represents a heavy quark. 

% for qft2006
%In the AdS/CFT framework, a heavy quark may be realized by a fundamental string which stretches from the asymptotic infinity (or from a ``flavor brane") to the black hole horizon. This string transforms as a fundamental representation; In this sense, the string represents a ``quark." The fundamental string has an extension and the tension, so the string has a large mass, which means that the string represents a heavy quark.

This kind of string has been widely studied in the past to measure the heavy quark potential \cite{Maldacena:1998im,Rey:1998ik}. For a $q\bar{q}$ pair, two individual strings extending to the boundary is not the lowest energy configuration. Instead, it is energetically favorable to have a single string that connects the pair (\fig{wilson}). The energy difference is interpreted as a $q\bar{q}$ potential, and one gets a Coulomb-like potential for ${\cal N}=4$.

At finite temperature \cite{Rey:1998bq,Brandhuber:1998bs}, it is no longer true that a string connecting the $q\bar{q}$ pair is always the lowest energy configuration; for large enough separation of the pair ($L_s$), isolated strings are favorable energetically. This phenomenon is the AdS/CFT description of the Debye screening and $L_s$ may be interpreted as the screening length.

References~\cite{Liu:2006nn,Chernicoff:2006hi,Peeters:2006iu} proposed how to compute the velocity dependence of the screening length.%
\footnote{See also Refs.~\cite{Caceres:2006ta}-\cite{Hoyos:2006gb} for extensions and the other issues.}
They computed the screening length in the $q\bar{q}$ rest frame, {\it i.e.}, they considered the plasma flowing at a velocity $v$. Such a ``plasma wind" is obtained by boosting a black hole.

At the leading order in $v$, the screening length is obtained as
%\cite{Liu:2006nn,Chernicoff:2006hi,Peeters:2006iu,Caceres:2006ta}
\bea
\lefteqn{(\mbox{screening length})} 
\nonumber\\
&\propto& \frac{1}{\epsilon_0^{1/4}} (1-v^2)^{1/4} \\
&\sim& (\mbox{boosted plasma energy density})^{-1/4}
\nonumber
\eea
for the ${\cal N}=4$ SYM, where $\epsilon_0$ is the unboosted energy density. Thus, the screening effect becomes stronger than the $v=0$ case. One can also compute the screening length for the other gauge theories \cite{Caceres:2006ta}. The leading behavior in $v$ seems universal. Namely, if one writes the screening length as
\be
(\mbox{screening length}) \propto (1-v^2)^\nu~,
\ee
the exponent $\nu$ is determined by the speed of sound $c_s$:
\be
4\nu = 1 - \frac{3}{4}(1-3c_s^2) + \cdots
\label{eq:mu_vs_cs}
\ee
when the theory is nearly conformal, {\it i.e.}, $c_s^2 \sim 1/3$. One can make a simple estimate of the exponent for QCD. According to the lattice results cited in Ref.~\cite{Karsch:2006sm}, all groups roughly predict $1/3-c_s^2 \sim 0.05$ around $2T_c$. Bearing in mind that our results are valid to large-$N$ theories and not to QCD, \eq{mu_vs_cs} 
gives $\nu \sim 0.22$. It would be interesting to compare this number with lattice 
calculations and experimental results.

One can also study the screening length at finite chemical potential \cite{Caceres:2006ta,Avramis:2006em}. At the leading order, the screening length at finite chemical potential is the same as the one at zero potential for a given energy density.

%%===========================================================
\subsection{Jet quenching}\label{sec:jet}
%%===========================================================

Another interesting QGP phenomenon is jet quenching. In the parton hard-scattering, jets are formed, but the jets have to travel in the QGP medium, so the jets are strongly suppressed. This phenomenon is known as jet quenching. So, the interesting quantity is the energy loss rate of partons. The AdS/CFT descriptions of jet quenching have been proposed recently \cite{Liu:2006ug}-\cite{Gubser:2006bz}. The proposals have been quickly extended to the other gauge theories \cite{Buchel:2006bv}-\cite{Nakano:2006js}.%
\footnote{See Refs.\cite{Friess:2006aw}-\cite{Argyres:2006yz} for the other issues. See also Ref.~\cite{Sin:2004yx} for an early attempt.}
%\footnote{For an extensive list of literature, see, {\it e.g.}, Ref.~\cite{Caceres:2006ta}.} %See also Ref.~\cite{Sin:2004yx} for an early attempt.}

To discuss jet quenching, one now moves the fundamental string with a  velocity $v$ along a brane direction. Then, the momentum carried by the string flows towards the horizon and one interprets the flow as the energy loss rate. For example, the energy loss rate for the ${\cal N}=4$ SYM becomes
\be
\frac{dp}{dt} = - \frac{\pi}{2} \sqrt{\lambda}T^2 \frac{v}{\sqrt{1-v^2}}~.
\ee
%where $\lambda := g_{\rm YM}^2 N$ is the `t~Hooft coupling.
%For example, the so-called ``friction coefficient" $\mu$ is defined by
%\be
%\frac{dp}{dt}=-\mu p
%\ee
%where $p$ is the momentum of the string.

% for hep-ph
Unfortunately, the result obtained in this way has some drawbacks. First, the result is not universal and is model-dependent. Second, the black hole results become exact only in the $\lambda\rightarrow\infty$ limit, but the result does not have a finite large-$\lambda$ limit. These two drawbacks are in contrast to the $\eta/s$ case. This may suggest that one has to be careful to apply AdS/CFT to QGP. Namely, one should not always take AdS/CFT results at face value.

% for qft2006
%Unfortunately, the result obtained in this way has some drawbacks. First, the result is not universal and is model-dependent. Second, the black hole results become exact only in the $\lambda\rightarrow\infty$ limit, but the result does not have a finite large-$\lambda$ limit. These two drawbacks are in contrast to the $\eta/s$ case. One can still try to put the numerical values naively, but the value obtained in this way is not close to the experimentally favored value. It is still too early to draw conclusions, but this may suggest that one has to be careful to apply AdS/CFT to QGP. Namely, we should focus on the universality, and naive extrapolations of ${\cal N}=4$ results may not be a good idea.

%%%%%%%%%
\section{Towards ``AdS/QGP"}
%%%%%%%%%

%\begin{sloppypar}
Hydrodynamic description of gauge theory plasmas using AdS/CFT is very powerful due to the universality. The AdS/CFT may be useful to analyze experiments. Conversely, experiments or the other theoretical approaches (such as lattice gauge theory) may be useful to confirm AdS/CFT. This approach is also important since currently there are many loose ends on the AdS/CFT derivation. One has to clear up these loose ends, but it may be hard to make further progress within string theory alone. So, the inputs from the other areas may be useful to increase confidence in the AdS/CFT derivation. On the other hand, one has to be careful to apply AdS/CFT to QGP if the universality does not hold ({\it e.g.}, jet quenching).
%\end{sloppypar}

In this talk, I emphasized the universality approach, but of course finding the gravity dual of QCD is desirable. The biggest assumption of the universality argument is that such a dual indeed exists, so finding the dual is necessary for the universality approach as well. One well-known model of QCD is the Sakai-Sugimoto model \cite{Sakai:2004cn}. This was a successful model in the confining phase, but unfortunately it is not a good model in the plasma phase. The Sakai-Sugimoto model involves the D4-brane compactified on a circle. Above the deconfinement temperature, the model looks as a true five-dimensional theory. 

Another problem of the current approach concerns the large-$\lambda$ limit; AdS/CFT and QGP are actually different limits. The large-$\lambda$ limit of the AdS/CFT is $g_{\rm YM} \rightarrow 0$ and $N \rightarrow\infty$. But of course QGP has a large-$\lambda$ since $g_{\rm YM} \sim O(1)$ and $N = 3$. We are not taking the same limits, so it is not clear why both give such close results.

%%%%%%%%%
%\section*{Acknowledgments}
%%%%%%%%%

% revtex
\begin{acknowledgments}
%\begin{sloppypar}
I would like to thank Elena C\'aceres, Kengo Maeda, and Takashi Okamura for collaboration and for their comments on the manuscript. It is a pleasure to thank Testuo Hatsuda, Tetsufumi Hirano, Kazunori Itakura, Tetsuo Matsui, Osamu Morimatsu, Berndt M\"{u}ller, and Tadakatsu Sakai for useful conversations. I would also like to thank the organizers of various workshops for the opportunity to give this lecture. The research of M.N.\ was supported in part by the Grant-in-Aid for Scientific Research (13135224) from the Ministry of Education, Culture, Sports, Science and Technology, Japan.
%I would like to thank Elena C\'aceres, Kengo Maeda, and Takashi Okamura for collaboration. It is a pleasure to thank Testuo Hatsuda, Tetsufumi Hirano, Kazunori Itakura, Tetsuo Matsui, Osamu Morimatsu, Berndt M\"{u}ller, and Tadakatsu Sakai for useful conversations. I would also like to thank the organizers of the YITP Workshop for the opportunity to give this lecture. The research of M.N.\ was supported in part by the Grant-in-Aid for Scientific Research (13135224) from the Ministry of Education, Culture, Sports, Science and Technology, Japan.
%\end{sloppypar}
\end{acknowledgments}

\appendix

%%%%%%%%%
\section{The AdS/CFT duality}\label{sec:ads-cft}
%%%%%%%%%

The precise correspondence between gauge theory and gravity is as follows:
\bea
\lefteqn{{\cal N}=4 \mbox{ super-Yang-Mills theory (SYM)} \leftrightarrow}
\nonumber\\
&& \mbox{type IIB string theory on AdS}_5 \times S^5.
\label{eq:dual2}
\eea
\noindent
One can reach this correspondence by studying the D3-brane more carefully, but instead I explain the correspondence from the symmetry point of view in this Appendix.

%%===========================================================
\subsection{Why ${\cal N}=4$ SYM}\label{sec:SYM}
%%===========================================================

To understand why the ${\cal N}=4$ SYM appears, let us go back to the open string oscillations on the D3-brane. 
Superstring theory actually requires 10-dimensional spacetime for consistency. The open strings on a D-brane are bounded on the D-brane, so the D3-brane represents a 4-dimensional gauge theory, but these open strings still oscillate in the full 10-dimensional spacetime. Thus, the simplest open string oscillations have 8 degrees of freedom instead of 2 discussed 
%YKIS
%above. 
in the text.
What are these degrees of freedom? In other words, what kind of gauge theory the D3-brane represents?

To see this, note that there are two types of string oscillations. First are the oscillations in the brane and the other are the oscillations out of the brane. From the 4-dimensional point of view (in terms of $SO(1,3)$ representations), the former represent a gauge field. The latter represent scalars. The spatial dimension is 9 and the brane dimension is 3, so there are 6 scalar fields. Thus, the gauge theory represented by the D3-brane inevitably comes with scalar fields. The ${\cal N}=4$ SYM is such a theory.
%YKIS
%Such a gauge theory is known as the ${\cal N}=4$ super-Yang-Mills theory (SYM).

The field contents of the ${\cal N}=4$ SYM include the gauge field $A_\mu$ and the scalars $\phi_i$. (The color indices are suppressed for simplicity. We will also suppress the spinor indices.) In addition, there are 4 fermions $\lambda_I$ due to supersymmetry (which comes from the supersymmetry in superstring.) The Lorentz transformation properties are different, but they come from similar string oscillations, which means that all these fields transform as the adjoint representation. Namely, the theory has no fundamental representation such as quarks. 
%YKIS
%How can one include fundamental matter? This is the topic of \sect{heavy_quark}, where I will explain a simple way to describe matter in the fundamental representation.

Also, if we have only D3-branes, as in the present case, the directions transverse to the brane are all isotropic. These directions correspond to the scalar fields, so the isotropy means that there is a global $SO(6)$ symmetry for $\phi_i$. Such a global symmetry is known as R-symmetry. 

The action for the ${\cal N}=4$ SYM is given by
\bea
{\cal L} &=& \frac{1}{g_{YM}^2}
\{ -\frac{1}{4} F_{\mu\nu}^2 - \frac{1}{2} (D_{\mu}\phi_i)^2 
- \frac{i}{2} \bar{\lambda}_I \gamma^{\mu}D_{\mu}\lambda^I
\nonumber \\
&& \mbox{} + O(\phi^4) + O(\lambda\lambda\phi) \}~.
\eea
First 3 terms are standard kinetic terms for the gauge field, the scalar fields, and the fermions. In addition, the action contains interaction terms which are written only schematically; $\phi^4$ interactions and Yukawa-like interactions.

The gauge field and the scalars have mass dimension 1 and the fermions have mass dimension $3/2$, so all terms in the action have mass dimension 4, which means that the theory has no dimensionful parameter and the theory is scale invariant. Actually, it is often the case in a relativistic field theory that the scale invariance and Poincar\'{e} symmetry $SO(1,3)$ combine into a larger symmetry, the conformal symmetry $SO(2,4)$. Thus, the ${\cal N}=4$ SYM has the global $SO(2,4) \times SO(6)_R$ symmetry. 

%%===========================================================
\subsection{Why ${\rm AdS}_5 \times S^5$}\label{sec:ads}
%%===========================================================

The ${\rm AdS}_5$ space is a spacetime with constant negative curvature. We are familiar with spaces with positive curvature, like the sphere, and have no problem visualizing them, but it is more difficult to visualize a space with negative curvature. A space with constant negative curvature is known as a hyperbolic space. One way to visualize a hyperbolic space is illustrated in famous woodcuts by Escher ({\it e.g.}, ``Circle Limit III"). It is not easy to visualize ${\rm AdS}_5$, but one can represent ${\rm AdS}_5$ as a hypersurface in a flat spacetime, which is useful for our purpose.

The ${\rm AdS}_5$ space is described by
\be
X_0^2 + X_5^2 - X_1^2 - \cdots - X_4^2 =l^2
\label{eq:embedding}
\ee
in a six-dimensional flat space%
%\footnote{Even though the space (\ref{eq:flat}) has ``two times," the AdS itself has the signature $(-,+,\cdots,+)$.}
\be
ds_6^2 = -dX_0^2 - dX_5^2 + dX_1^2 + \cdots + dX_4^2~.
\label{eq:flat}
\ee
[Even though the space (\ref{eq:flat}) has ``two times," the AdS itself has the signature $(-,+,\cdots,+)$.] 
Here, $l$ gives the characteristic length scale of the ${\rm AdS}_5$ space and it is related to the `t~Hooft coupling from the discussion in the text.
%YKIS
%above. 
Note that if all signs are positive in \eq{embedding}, \eq{embedding} defines a sphere. Also, \eq{embedding} is a higher dimensional generalization of the standard hyperbola $x^2-y^2=1$. In this form, it is clear that ${\rm AdS}_5$ has a $SO(2,4)$ symmetry just like the ${\cal N}=4$ conformal symmetry; to say that theories are dual to each other, they should have the same symmetries.

In addition, the full geometry in \eq{dual2} involves $S^5$ which has a $SO(6)$ symmetry. This is the same symmetry as the ${\cal N}=4$ R-symmetry. So, the gravity side also has the global $SO(2,4) \times SO(6)$ symmetry. 

%\footnotesize

\end{document}